\documentclass[pra,showpacs,twocolumn]{revtex4}
\usepackage{graphicx}
\usepackage{amssymb}
\usepackage{color}
\usepackage{rotating}

\newcommand{\one}{\mbox{ \hspace{-0.5mm}1\hspace{-0.1cm}l}}
\newcommand{\dd}{\mbox{d}}
\newcommand{\half}{\frac{1}{2}}
\newcommand{\nn}{\nonumber} 
\newcommand{\Eqref}[1]{Eq.~(\ref{#1})}
\newcommand{\ket}[1]{\left|{#1}\right\rangle}
\newcommand{\bra}[1]{\left\langle{#1}\right|}
\newcommand{\scalar}[2]{\left.\left\langle{#1}\right|{#2}\right\rangle}

\definecolor{purple}{rgb}{0.85,0,0.85}

\begin{document}

\title{Adiabatic information transport in the presence of decoherence}

\author{I. Kamleitner} \author{J. Cresser} \author{J. Twamley}
\affiliation{Centre for Quantum Computer Technology, Physics Department, Macquarie University, Sydney, New South Wales 2109, Australia}

\begin{abstract} 
	We study adiabatic population transfer between discrete positions. Being closely related to \mbox{STIRAP} in optical systems, this transport is coherent and robust against variations of experimental parameters. Thanks to these properties the scheme is a promising candidate for transport of quantum information in quantum computing. We study the effects of spatially registered noise sources on the quantum transport and in particular model Markovian decoherence via non-local coupling to nearby quantum point contacts which serve as information readouts. We find that the rate of decoherence experienced by a spatial superposition initially grows with spatial separation but surprisingly then plateaus. In addition we include non-Markovian effects due to couplings to nearby two level systems and  we find that although the population transport exhibits robustness in the presence of both types of noise sources, the transport of a spatial superposition exhibits severe fragility.
\end{abstract}
\maketitle

\section{Introduction}

The coherent transport of quantum information is an essential element in any scalable architecture for a quantum information processor. The interconversion of static qubits into flying qubits is difficult and may not be a feasible solution for the coherent spatial transport of quantum information. As a possible alternative much attention has been focused on dark state adiabatic passage for coherent state quantum transport. Originally studied in the context of quantum optics \cite{overview} where it is called stimulated Raman adiabatic passage (STIRAP), dark state transport uses the existence of a ``dark state'' which is a zero-energy eigenstate of a driven quantum system. By manipulating the driving of the system one can sculpt this dark state to coherently transport quantum states using STIRAP-like procedures. This intra-atomic dark-state transport has been demonstrated experimentally \cite{Goto2006}. However the method has more recently been applied to  spatial transport of quantum information (which we specifically denote CTAP - coherent transport by adiabatic passage following \cite{main}),  in a variety of physical systems including chains of neutral atoms \cite{Eck}, quantum dots \cite{main, Hohenster2006, Petrosyan2006}, superconductors \cite{Siewert2006}, and photons in nearby waveguides \cite{Longhi2007}.  It has also been proposed as a crucial element in the scale up to large quantum processors \cite{Hollenberg}. The method possesses two very crucial benefits over other quantum transport methods: since the transport is via a zero energy state the quantum state acquires no dynamical phase and due to the adiabatic theorem, the process is very robust to a wide range of system variations. However, an important question, particularly with regard to the use of CTAP within large scale quantum computer architectures,  is to determine the effects of decoherence on the transport. The effects of dephasing and spontaneous emission has previously been examined in the case of STIRAP dark-state transport in a three level atom in a $\Lambda$ configuration \cite{Ivanov2004, Ivanov2005}. In that work a master equation was postulated and its effects  on the transport studied. 

In our work we examine the effects of two types of physically-motivated decoherence sources effecting the CTAP transport in a quantum chain. We first study the effects of delocalised measurements on the systems making up the CTAP chain. In particular we imagine an electron on a chain of quantum dots (QD)  and each is measured by a nearby quantum point contact (QPC). These QPCs however, are non-local measurement devices in that their charge sensitivity falls off with distance. Such devices or similar will be required to either modulate or readout the quantum information in each quantum dot in the CTAP chain in a real device. In large scale quantum processors one will routinely wish to have quantum information in a superposition of two ``distant'' spatial locations. It is known that from numerous studies of  cat-states in quantum Brownian motion - a single harmonic oscillator coupled to a bath of harmonic oscillators - the rate of decoherence suffered by the cat grows quadratically with the spatial separation of the two superposition states of the ``cat''. We find that such an effect is also present in our case, i.e. the decoherence rate of a ``cat-state'' on the CTAP chain increases with cat-seperation, but surprisingly we find that this decoherence rate saturates beyond a critical cat spatial seperation.  This is a positive result for the CTAP transport protocol and is essentially due to the rapid spatial fall off of the measurement sensitivity of the QPCs. 

This first model is an example of Markovian decoherence. We also include a second non-Markovian dephasing source, and consider that each quantum dot interacts with a nearby two level system (TLS). This could model unlocated two-level fluctuators in a solid state CTAP scheme.  Interestingly we find that qubit transport still seems relatively robust in the presence of these combined Markovian and non-Markovian decoherence sources. More worryingly however we find that this non-Markovian dephasing slightly entangles the two level systems with the transported qubit. This effect does not seriously detract from the transport of an electron isolated on one QD but causes serious degradation if the qubit transported is in a superposition state. As 
the latter situation will be the typical case in a large scale quantum processor the present analysis might indicate a much lower density of  unlocated TLSs will be required when using CTAP in large scale quantum computer.

We proceed in section \ref{model} to motivate and model two sources of decoherence and solve the corresponding master equation without interactions between different QDs.  Transport of the electron by CTAP is described in section~\ref{transport} under both the Markovian and non-Markovian noise models. Interestingly we are able to analytically obtain an equation describing the fidelity of the Markovian dynamics while we numerically solve the non-Markovian case.  It is here where differences between Markovian and non-Markovian dephasing become most apparent. We summarize our results and discuss their relevance in section~\ref{conclusion}.

\section{Model\label{model}}
\subsection{Measurements}
	\begin{figure}
		\centering
		\includegraphics[width=8.5cm]{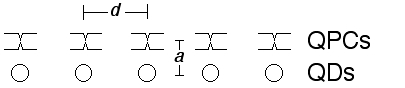}
		\caption{Realization of non-local measurement as proposed in \cite{QPC}. The electric current through a QPC is influenced by the appearance of an electronic charge in its near environment.  \label{fig1}}
	\end{figure}
	As we mention above, we first consider each of the CTAP quantum systems to be continually measured. This gives rise to Markovian decoherence. We take the specific example of an electron on a chain of quantum dots and consider the measurements to be made by quantum point contacts (QPC) situated close to the quantum dots as shown in Fig~\ref{fig1}. The measurement executed by a QPC is caused by the modulation of the current transmitted through a QPC  due to the presence of a nearby electron. The QPC current is modulated by a factor $1-\frac{\alpha}{r}$, where $r$ is the distance between electron and QPC {\color{black} and $\alpha$ is a constant reflecting the properties of the QPCs}  (see~\cite{QPC}). This modulation results in an indirect position measurement of the electron's spatial position on the  rail of QDs.  However, it is a non-local measurement because even an electron on the neighboring QD influences the transmission through a QPC. The localness is parameterized by $\frac{a}{d}$, i.e. the distance between QPC rail and QD rail over the distance between two neighboring QDs, with small values representing more local measurements. Furthermore $\frac{\alpha}{a}$ parameterizes the sensitivity (signal over noise) of the measurements and is typically small and hence the measurements are weak ones. Such measurements are properly described in the language of positive operator valued measurements (POVM)~\cite{lectures,Busch}.
	
	The purpose of a measurement apparatus is the readout of quantum information for which one would like to use strong local measurements. This can be approximated by using a large number of weak, non-local measurements of the type described. A large number of measurements in a reasonably short time is achieved by  using a high measurement rate which in turn can be realized, for instance, by applying a voltage to the QPCs.  However, such measurements also act as a source of decoherence. During quantum unitary operations such as transportation by CTAP it is preferred that this decoherence is absent. However this might not be completely achievable in practice and one may be left with a small decoherence rate due to the coupling to the QPCs during any quantum operation. Thus measurements as a source of decoherence are included in the analysis presented here.
	
	We restrict our treatment to the case of having only one electron  in the rail of QDs. We also assume that the electron can only occupy the ground state of the QDs, and we take $\ket{i}_{\!A}$, to be the quantum state of the electron in the $i^{th}$ QD. Furthermore we neglect all interactions depending on the spin of the electron. Then, $\{ \ket{i}_{\!A},\; i=1,..,N\}$, form a basis for the Hilbert space $\mathcal{H}_A$ of this electron on the QD rail with $N$ dots. In the following we take the limit of a long rail, $N\to\infty$. We denote the distance between the $i$-th QPC and the $j$-th QD by 
	\begin{equation}
		r_{ij}=\sqrt{a^2+(|i-j|d)^2} \label{distance}
	\end{equation}
	where $a$ and $d$ are defined in Fig~\ref{fig1}.

	The probability of the $j^{th}$ QPC detecting the presence of an electron on the QD rail  can be written as~\cite{lectures}
	\begin{equation}
		P_j(\rho^A) = \mbox{Tr}(\pi_j\rho^A) \label{prob}
	\end{equation}
	where $\rho^A$ is the state of the electron on the QD rail and $\pi_j$ is the effect operator corresponding to the QPC measurement at site~$j$. If the electron is spatially localised to be only on the $i^{th}$ QD, i.e. in the state $\ket{i}_A$, \Eqref{prob} reduces to 
	\begin{equation}
		P_j(\ket{i}_A\bra{i}) =  \hspace{3mm}\bra{i}_{\hspace{-6mm}A}\hspace{4mm}\pi_j\ket{i}_A.
	\end{equation}
	As we noted above the measurement sensitivity of the QPC depends on the distance $r_{ij}$. The presence of an electron a distance $r_{ij}$ away from the QPC decreases the current flowing through the QPC by a factor  $1-\frac{\alpha}{r_{ij}}$  and this leads to a reduced detection probability,
	\begin{equation}
		P_j(\ket{i}_A\bra{i}) \propto 1-\frac{\alpha}{r_{ij}}\;\;. \label{relprob}
	\end{equation}
 Fulfillment of \Eqref{relprob} is certainly achieved with the measurement effect operators
	\begin{equation}
		\pi_j = \frac{\gamma}{N} \sum_{i=1}^N \left(1-\frac{\alpha}{r_{ij}}\right)\ket{i}_{\!A}\!\bra{i} \label{effect}
	\end{equation}
	with
	\begin{equation}
		\gamma = \frac{N}{ \sum_{j=1}^N \left(1-\frac{\alpha}{r_{ij}}\right)} \parbox{1.4cm}{{\small \hspace{2mm}$N\!\to\! \infty$}\\\phantom{,}  \quad$=$\\{\small \phantom{hallo}}} 1 + \sum_{j=1}^{N}\frac{\alpha}{Nr_{ij}}
	\end{equation}
	to satisfy $\sum_{j=1}^{N}\pi_j=\one$. Note that each effect operator is almost proportional to the unit operator which reflects the weakness  of the measurements being performed.

	Measurement theory states that the transformation of the density operator due to a measurement  is described by \cite{lectures, Busch}
	\begin{equation}
		\rho^A \quad\parbox{1.2cm}{{\small $\;\;\pi_j$}\\$\overrightarrow{\phantom{hihh}}$} A_j \rho^A A_j^\dagger
	\end{equation}
		with $A_j=U_j\sqrt{\pi_j}$ for some arbitrary unitary operators $U_j$. This unitary depends on the interaction of the quantum system and measurement apparatus during the measurement process. In our case, if the electron is at site $i$, its position will not be changed by a detection event by a QPC at site $j$. For simplicity we also assume that the measurement does not introduce relative phases within the rail of QDs giving $U_j\equiv \one$ and $A=A^\dagger$, which means the measurements influence the electron state as little as possible.
	
	To derive the master equation we now assume that detection events in the QPCs occur uniformly at random and at a constant rate $R$. Following \cite{MME}, we can write down the master equation describing the evolution of the density operator as
	\begin{equation}
		\frac{\dd \rho^A}{\dd t} = -\imath[H_A,\rho^A] - R\rho^A + R\sum_{j=1}^N A_j \rho^A A_j .\label{4}
	\end{equation}
	This equation is in Lindblad form with Lindblad operators $A_j$ and thus the evolution is Markovian. 
	For the case when $H_A$ involves no electron transport along the QD rail, Eqn. \Eqref{4} possesses a stationary state which is diagonal in $\ket{i}_A$, and thus the evolution corresponds to a pure dephasing type of decoherence. We note that $R$ should scale proportional to $N$ as each QPC contributes equally to the overall measurement rate.  
	
	For now we take $H_A=0$, but later we introduce a time dependent Hamiltonian to induce CTAP. Expressing \Eqref{4} in the basis $\ket{k}_A$,  we obtain $\dot{\rho}^A_{kk}=0$, for diagonal entries and $\dot{\rho}^A_{kl}=-T_{kl}\rho^A_{kl}$ for off-diagonal ones. The decoherence rate
	\begin{equation}
		T_{kl} = R\left[-1+\frac{\gamma}{N}\sum_{j=1}^N \sqrt{1-\frac{\alpha}{r_{kj}}} \sqrt{1-\frac{\alpha}{r_{lj}}}\right] \label{5}
	\end{equation}
	($r_{ij}$ is defined in~\Eqref{distance}) can be shown to saturate for large distances \mbox{$|k-l|d\gg a$}. To simplify these formulas one can use the limit of non-sensitive measurements $\frac{\alpha}{r_{ij}}\ll 1$ which should be well justified in experiments~\cite{QPC}. This weak measurement limit corresponds with the inability of the measurements to give detailed information on the position of the electron in the QD rail. Weak measurements of this nature feature in many models of continuous monitoring of a quantum particle's position \cite{Milburn-Caves}.  In the limit of weak measurements we obtain
	\begin{equation}
		T_{kl} \approx \frac{R\alpha^2}{8N}\sum_{j=1}^N \left( \frac{1}{r_{kj}}-\frac{1}{r_{lj}} \right)^2
	\end{equation}
 	which, for spatial separations larger than the threshold,  \mbox{$|k-l|d\gg a$}, limits to 
	\begin{equation}
		\frac{R\alpha^2}{4Nd^2}\frac{\coth{(\pi a/d)}}{a/d}. \label{coth}
	\end{equation}
	
We also note that we can execute local measurements, which give maximal information about the particle's position in the limit $\frac{1}{r_{ij}}=\delta_{ij}$, while keeping $\alpha$ finite, e.g. in \Eqref{5}. In this case we find 
	\begin{equation}
		\color{black}T_{kl}=\frac{R}{N}\left( 2+\alpha-2\sqrt{1+\alpha}\right).\label{aaaa}
	\end{equation}
	
	The saturation of the decoherence rate in \Eqref{coth}, is somewhat surprising when one compares this with the similar situation for a free particle in a spatial superposition cat-state, experiencing continuous position measurements~\cite{Joos}.  In that case the decoherence rate suffered by the particle increases without bound according to the spatial separation of the cat, i.e. $T_{x_1,x_2}\propto (x_1-x_2)^2$. 

\subsection{Coupling to TLSs\label{coupling1}}
	\begin{figure}
		\centering
		\includegraphics[width=8cm]{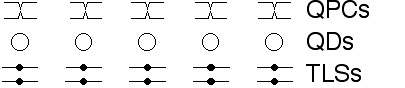}
		\caption{As in Fig~\ref{fig1} but with additional local coupling to two-level systems. \label{fig2}}
	\end{figure}

We now consider a further source of decoherence. It is highly likely that in any physical device there will be unknown accidental two level fluctuators nearby to the quantum dot rail (see Fig. \ref{fig2}). If these unknown two level systems (TLS), can couple to the electron on the rail then these  systems effect a source of decoherence which exhibits memory, i.e. is non-Markovian. Quantum coherences on the quantum dot rail can be transferred to the nearby TLSs, where they can remain for a period, before being transferred back. Typically the analysis of these types of non-Markovian effects are complex but in the following we are able to derive analytic solutions of the resulting reduced dynamics of the quantum dot rail. For simplicity we assume these fluctuators have no internal dynamics other than their coupling to the quantum dot rail \footnote{A possible internal Hamiltonian $H_{TLS}$ does not influence the QD rail if $[H_{TLS},H_{int}]=0$, because it can be removed by a unitary operation acting on the TLS-Hilbert space only.}. Furthermore we assume that these TLSs do not experience significant decoherence on the time scales of the transport. We are aware that these assumptions may not be completely satisfied in all realistic situations. However this model will serve to highlight the striking difference between the Markovian and non-Markovian evolutions in, for instance, the transport of a spatial  superposition.
	
	We use $\ket{1}_{\!j}$ and $\ket{0}_{\!j}$ as basis of the Hilbert space $\mathcal{H}_j$ of the $j-$th TLS such that the interaction Hamiltonian is diagonal. If the electron is on the $j$-th QD, it is assumed to induce a phase shift on the $j-$th TLS, so that the Hamiltonian acting in the {\color{black}product Hilbert space $\mathcal{H}=\mathcal{H}_A\bigotimes_{j=1}^N \mathcal{H}_j$ of electron and TLSs} reads
	\begin{equation}
		H_{int} = \sum_{n=1}^{N}\,\left[ \chi_n\ket{n}_{\!A}\!\bra{n}\otimes\sigma_{z,n}\bigotimes_{j\ne n} \one_j\right]\;\;. \label{inter}
	\end{equation}
	{\color{black}The coupling constants $\chi_j$ are} considered to be constant in time, and $\sigma_{z,j}$ and $\one_j$ are the Pauli $Z$-matrix and the identity operator, respectively, acting in $\mathcal{H}_j$. 
	
	As typically assumed, we now take the initial state to be in product form $\rho(t=0)=\rho_0^A\otimes\rho_{0}^{TLSs}$, where $\rho_{0}^{TLSs}$ does not have to be a product state of the different TLSs. We can now express the master equation describing the time evolution of the density matrix of the combined QD rail and TLSs under the effects of the above measurements  (see \Eqref{4}), to be 
	\begin{equation}
		\frac{\dd \rho}{\dd t} = -\imath[H_{int},\rho] - R\rho + R\sum_{j=1}^N A_j \rho A_j \label{8}
	\end{equation}
	with $A_j=\sqrt{\pi_j}\bigotimes \one^{\otimes\,N}$.
		
	After some effort, one can trace out the TLSs to find the non-Markovian master equation for the reduced density matrix of the QD rail for an arbitrary initial product state of the TLSs, as
	\begin{eqnarray}
		\frac{\dd \rho^A}{\dd t} &=&  - R\rho^A+\sum_{n=1}^N \Big(R A_n\rho^AA_n - \imath\Delta_n(t) \left[\ket{n}_{\!A}\!\bra{n},\rho^A\right]  \nn\\
		& & +  \gamma_n(t) \big[  \ket{n}_{\!A}\!\bra{n}, \left[ \rho^A,\ket{n}_{\!A}\!\bra{n} \right]\big]  \Big) ,\label{stuff1}
	\end{eqnarray}
	where the last two terms describe the effects of the TLSs on the QD rail.
	Here the definition 
	\begin{equation}
		\gamma_n(t)-\imath\Delta_n(t) = \chi_n\frac{\sin{\chi_nt}-\imath\omega_n\cos{\chi_nt}}{\cos{\chi_nt}+\imath\omega_n\sin{\chi_nt}} 
	\end{equation}
	is used and $\omega_n=$Tr$[\rho_n\sigma_{z,n}]$, is the inversion of the $n$-th TLS. Decoherence due to the TLSs is described by $\gamma_n$ whereas $\Delta_n$ represents the Lamb-shift. If $\omega_n=\pm 1$, then $\Delta_n=\pm\chi_n$ and $\gamma_n=0$. In this case the decoherence due to the coupling to TLSs vanishes and the coupling to the TLSs results only in a change of the energy of states $\ket{n}_A$ by $\pm\chi_n$. However, a more interesting case is when $\omega_n=0$, since it includes the case where the TLSs may initially be in a complete mixture $\rho_n(0)=\half\one$.  In  this case we find $\Delta_n=0$, and $\gamma_n=\chi_n\tan{\chi_nt}$. The presence of singularities in $\gamma_n(t)$, may cause difficulties in studying the time evolution of $\rho^A$, using normal methods. To avoid this we do not work directly with (\ref{stuff1}). Instead we return to solve the complete dynamics of the coupled QDs and TLSs and then trace out the latter to obtain $\rho^A(t)$.

		To achieve this we first we go to the interaction picture by transforming \Eqref{8} with
	\begin{equation}
		e^{\imath H_{int}t} = \sum_{n=1}^{N} \ket{n}_{\!A}\!\bra{n}\otimes \left(\cos{\!(\chi_nt)}\!\one_n +\imath \sin{\!(\chi_nt)} \sigma_{z,n}\right) \bigotimes_{j\ne n} \!\one_j .
	\end{equation}
	The combined density operator in this picture is given by  $\overline{\rho}(t)=e^{\imath H_{int}t}\rho(t)e^{-\imath H_{int}t}$, and is governed by the master equation
	\begin{equation}
		\frac{\dd \overline{\rho}}{\dd t} =  - R\overline{\rho} + R\sum_{j=1}^N A_j \overline{\rho} A_j. \label{10}
	\end{equation}
	Note that $e^{\imath H_{int}t}A_je^{-\imath H_{int}t}=A_j$. Taking the components of \Eqref{10} in the QD Hilbert space,  $\overline{\rho}_{kl}(t)={}_A\!\bra{k}\overline{\rho}\ket{l}_{\!A}$ (which is still an operator in the TLSs Hilbert space), we find the solution is given in a similar manner to that of the previous subsection to be 
	\begin{equation}
		\overline{\rho}_{kl}(t) = e^{-T_{kl}t} \overline{\rho}_{kl}(0)= e^{-T_{kl}t} {\rho}_{kl}(0)
	\end{equation}
	where  $T_{kl}$ is given in \Eqref{5}. After transforming back to the Schr\"odinger picture and tracing over the TLSs we find for the components of the reduced density operator 
	\begin{eqnarray}
		\rho^A_{kk}(t) &=& \rho^A_{kk}(0) \\
		\rho^A_{kl}(t) &=& e^{-T_{kl}t} \cos{(\chi_kt)}\cos{(\chi_lt)}\rho^A_{kl}(0) \label{13}
	\end{eqnarray}
	where we have assumed that the initial states of the TLSs is a completely mixed state
	\begin{equation}
		\rho_j(0)=\half\one . \label{initial}
	\end{equation}
	Hence we see that information lost to the TLSs can return to the rail via the oscillatory terms in~\Eqref{13}, which is in contrast to Markovian decoherence induced by the measurements. Note that the non-Markovian behavior of~\Eqref{13} results from tracing out the environmental TLSs.

\section{CTAP\label{transport}}

	Coherent Tunneling by Adiabatic Passage relies on the adiabatic theorem in the sense that the electron will always be in an eigenstate of the time dependent Hamiltonian. For the transport from the $m$-th to the $n$-th QD, the Hamiltonian is designed such that there exists an eigenstate which is $\ket{m}_A$ at the beginning of the transport, and then changes continuously to $\ket{n}_A$ during transport. This way of population transfer has the important property of being robust against small variations of the Hamiltonian and the transport time. This is crucial in many experiments since these parameters are often hard to control. The drawback of CTAP is a relatively long transport time which is limited by the adiabatic theorem and usually is about an order of magnitude longer compared to diabatic population transfer.

	To transport an electron from position $m$ to position $n$ we assume that we can control tunneling rates between neighbouring QDs, but we do not require any next to neighbour coupling. The  Hamiltonian then reads	
	\begin{equation}
		\hspace{3mm}H_A = \left( \begin{array}{ccccccc}
			  0 & \Omega_m&0& \cdots& 0&0 &0 \\ \vspace{0mm}
			  \Omega_m & 0 &\!\!\!\Omega_{m+1} \!\!\! & &0& 0&0 \\ 
			  0 &\!\!\! \Omega_{m+1} \!\!\! & 0 & &0&0&0 \\
			  \vdots&  & &\ddots  &&&\vdots \\
			  0&0&0&&0&\!\!\!\Omega_{n-2}&0 \\
			  0 &0&0 && \!\!\!\Omega_{n-2} &0& \!\!\!\Omega_{n-1}  \!\!\! \\
			  0 & 0 &0&\cdots&0&  \!\!\!\Omega_{n-1}  \!\!\!& 0  \\
		\end{array} \right). \label{14}
	\end{equation}

	The time dependent manipulation of the tunneling rates $\Omega_j$ can be achieved in an experimental setup via external gates as is outlined in~\cite{main}. The tunneling rates $\Omega_m=\Omega_P$ and $\Omega_{n-1}=\Omega_S$ are often called pump and Stokes pulse, respectively. The application of CTAP can roughly be outlined as follows:
	\begin{enumerate}
		\item While keeping $\Omega_m=0$ one switches $\Omega_j=\Omega_{max}$ for $j=m+1,\dots,n-1$ to ensure all energy eigenstates for an electron situated between $m$ and $n$ are non-degenerate. 
		\item Then $\Omega_m$ is increased and $\Omega_{n-1}$ is decreased until it vanishes. This process has to be done slowly to satisfy adiabaticity. In this step the electron moves from the $m-$th to the $n-$th QD.
		\item Finally all couplings can be set to zero.
	\end{enumerate}
	To avoid any geometric phases one can use real and positive tunneling rates $\Omega_j$. In the following we will refer to these steps  as step one, step two and step three. {\color{black}We emphasize that the electron moves exclusively in the second step and therefore} the first and  the third steps can be done arbitrarily fast (up to experimental limitations). Hence we set $t=t_0$ at the beginning of step two and $t=t_{max}$ at the end of step two. It will be assumed that at $t=t_0$ the probability of finding the electron on the $j$-th QD with $j=m+1,\cdots,n$ is zero. This is reasonable since the adiabatic theorem does not hold for these states and we want to transport the electron from the $m$-th QD. Furthermore we use that $n-m$ is even, because only then all states are non-degenerate after step one.
	
	\subsection{Measurements\label{hello}}
	To introduce our technique for solving the master equation we first neglect the coupling to the TLSs.  At $t=t_0$ the $(n-m+1)$ eigenstates of $H_A$ which are super positions of $\ket{m}_A,\cdots,\ket{n}_A$, are denoted by $\ket{\psi_j(t_0)}$ with $j=\frac{m-n}{2},\cdots,\frac{n-m}{2}$, ordered by their energy $\varepsilon_j(t_0)$. These eigenstates evolve continuously to $\ket{\psi_j(t)}$. The un-normalized adiabatic state responsible for the transport reads~\cite{main}
	\begin{eqnarray}
		\ket{\psi_0(t)} &=& \cos{\Theta}\ket{m}_A+(-1)^{\frac{n-m}{2}}\sin{\Theta}\ket{n}_A \nn\\
		& & -X \sum_{j=2}^{\frac{n-m}{2}}(-1)^j\ket{2j-2+m}_A \label{psi0}
	\end{eqnarray}
	with
	\begin{eqnarray*}
		\Theta = \arctan{\frac{\Omega_m}{\Omega_{n-1}}} ,&\;& X= \frac{\Omega_m\Omega_{n-1}}{\Omega_{max}\sqrt{\Omega_m^2+\Omega_{n-1}^2}}  .
	\end{eqnarray*}

	Note that $\ket{\psi_0(t_0)}=\ket{m}_A$ and $\ket{\psi_0(t_{max})}=\ket{n}_A$, i.e. the states in which the electron is on the $m^{th}$ and $n^{th}$ QD, respectively. Furthermore during the entire process $\varepsilon_0(t)=0$ holds, which ensures that no dynamic phase appears for the state to be transported. If 
	\begin{equation}
		|\varepsilon_0(t)-\varepsilon_j(t)| \gg\left|\!\scalar{\frac{\dd}{\dd t}\psi_0(t)}{\psi_j(t)}\!\right|\quad \label{15}
	\end{equation}
	 holds and if the system at $t=t_0$ is in $\ket{\psi_0(t_0)}$, then the adiabatic theorem states that the system will stay in $\ket{\psi_0(t)}$, provided it is a closed system. Therefore an electron starting in $\ket{n}_A$ will end up in $\ket{m}_A$. 
	 
	 To generalize this concept to the open system described here, we follow~\cite{adiabat} and transform \Eqref{4} with the unitary operators defined by
	\begin{equation}
		\hspace{-1mm}\begin{array}{rcll}
			U^\dagger(t)\ket{\psi_j(t)} &\!\!=\!\!& \ket{\psi_j(t_0)}&\;\mbox{for }j=\frac{m-n}{2},\cdots,\frac{n-m}{2} \\
			U^\dagger(t)\ket{j}_A &\!\!=\!\!& \ket{j}_A & \;\mbox{for } j< m \mbox{ and }j> n
		\end{array}\hspace{-1mm} \label{uni}
	\end{equation}
	to get
	\begin{eqnarray}
		 \frac{\dd\tilde{\rho}^A}{\dd t} &=& -\imath\left[ \sum_{j=\frac{m-n}{2}}^{\frac{n-m}{2}} \varepsilon_j(t)\ket{\psi_j(t_0)}\!\bra{\psi_j(t_0)} - \imath U^\dagger\frac{\dd U}{\dd t} , \tilde{\rho}^A \right] \nn \\
		& & - R\tilde{\rho}^A + R\sum_{i=1}^N \widetilde{A}_i \tilde{\rho}^A \widetilde{A}_i \label{label}
	\end{eqnarray}
	with $\widetilde{O}= U^\dagger O U$ for an operator $O$. If, in addition to \Eqref{15} we also assume weak coupling to the environment \footnote{This is usually justified in systems described by a Markovian master equation.}, one can neglect the term $\imath U^\dagger\frac{\textrm{\footnotesize d} U}{\textrm{\footnotesize d}t}$ in \Eqref{label} 
as is shown in~\cite{adiabat}. This is the generalization of the adiabatic theorem to systems described by a master equation of Lindblad form. Hence we have achieved the time independence of the eigenspaces of the transformed Hamiltonian $\widetilde{H}$:
	\begin{eqnarray}
		\hspace{-0mm} \frac{\dd\tilde{\rho}^A}{\dd t} &=& -\imath\left[ \sum_{j=\frac{m-n}{2}}^{\frac{n-m}{2}} \varepsilon_j\ket{\psi_j(t_0)}\!\bra{\psi_j(t_0)} , \tilde{\rho}^A \right] \nn \\
		& & - R\tilde{\rho}^A + R\sum_{i=1}^N \widetilde{A}_i \tilde{\rho}^A \widetilde{A}_i \label{17}.
	\end{eqnarray}

	For $R=0$ we get the von Neumann equation
	\begin{equation}
		\frac{\dd\tilde{\rho}^A}{\dd t} =-\imath\left[ \sum_{j=\frac{m-n}{2}}^{\frac{n-m}{2}} \varepsilon_j\ket{\psi_j(t_0)}\!\bra{\psi_j(t_0)} , \tilde{\rho}^A \right] 
	\end{equation}
 and $\tilde{\rho}^A_{00}(t):=\bra{\psi_0(t_0)}\tilde{\rho}^A(t)\ket{\psi_0(t_0)}= {}_A\!\bra{m}\rho^A(t_0)\ket{m}_A$ is therefore a constant of motion, which ensures perfect transport of the electron.  At $t=t_{max}$ we use $U(t_{max})$ to transform back to the Schr\"odinger picture. From \Eqref{uni} we find again that an electron initially in state $\ket{m} _A =\ket{\psi_0(t_0)}  $ will be  in state $\ket{\psi_0(t_{max})}=\ket{n} _A $ at the end of step two.
 
 For $R\ne 0$ we can use \Eqref{17} to calculate the loss from perfect transport
	\begin{eqnarray}
		\hspace{-1mm} \frac{\dd}{\dd t}\tilde{\rho}^A_{00}(t) &=& - R\tilde{\rho}^A_{00}(t) + R\sum_{j=1}^N \bra{\psi_0(t)} A_j \rho^A(t) A_j \ket{\psi_0(t)}  \nn\\
		& = & - R\tilde{\rho}^A_{00}(t) +\frac{R\gamma}{N}\sum_{j=1}^N\sum_{i,i'=m}^{n} \sqrt{1-\frac{\alpha}{r_{ij}}} \nn\\ 
		& & \times\sqrt{1-\frac{\alpha}{r_{i'j}}}  \scalar{\psi_0(t)}{i}_{\!A}\!\! \scalar{i'}{\psi_0(t)}  \rho^A_{ii'}(t) \label{18}
	\end{eqnarray}
	As we argued before, without measurements $\tilde{\rho}^A(t)$ is a constant in time in the adiabatic approximation. To first order in the measurements \footnote{That is for small measurement rates $\frac{R}{N}$ and/or for weak measurements, i.e. small $\frac{\alpha}{a}$.} we can therefore substitute $\tilde{\rho}^A(t)\to \tilde{\rho}^A(t_0)=\ket{\psi_0(t_0)}\!\bra{\psi_0(t_0)}$ and $\rho^A(t)\to \ket{\psi_0(t)}\!\bra{\psi_0(t)}$ on the right side to find

	\begin{eqnarray}
		\hspace{-0.5mm}\frac{\dd}{\dd t}\tilde{\rho}^A_{00}(t) \!&\approx&\! -R\Bigg(\!1 - \frac{\gamma}{N}\sum_{j=1}^N \!\left[\sum_{i=m}^{n} \sqrt{1-\frac{\alpha}{r_{ij}}}  \left| \!\scalar{\psi_0(t)}{i}_{\!A}\right|^2 \!\right]^{\!2}  \!\Bigg) \nn\\&&\label{19}
	\end{eqnarray}
	 where $\ket{\psi_0(t)}$ can be substituted from \Eqref{psi0} to obtain numerical values. 
	 
	 One might also ask what happens to coherences $\tilde{\rho}^A_{k0}(t):= {}_A\!\bra{k}\tilde{\rho}^A(t)\ket{\psi_0(t_0)}$ (for any $k<m$ or $k>n$)  during the transport. This question arises if one is interested in the transport of one part of a super position state, e.g. $\frac{1}{\sqrt{2}} \left(\ket{k}_A+\ket{m} _A\right)\to\frac{1}{\sqrt{2}}  (\ket{k}_A +\ket{n} _A)$. In the same manner as \Eqref{18} and \Eqref{19} we arrive at
	\begin{eqnarray}
		\hspace{-4mm} \frac{\dd}{\dd t}\tilde{\rho}^A_{k0}(t) &=& - R\tilde{\rho}^A_{k0}(t) + \frac{R\gamma}{N}\sum_{j=1}^N  \sqrt{1-\frac{\alpha}{r_{kj}}} \nn\\
		& &  \times\sum_{i=m}^{n} \sqrt{1-\frac{\alpha}{r_{ij}}} {}_A\!\!\scalar{i}{\psi_0(t)}\rho_{ki}(t) \label{20} \\
		&\approx& -R\tilde{\rho}_{k0}^A(t_0) \Bigg( 1-\frac{\gamma}{N} \sum_{j=1}^N  \sqrt{1-\frac{\alpha}{r_{kj}}}\nn\\
		& & \times \sum_{i=m}^{n}  \sqrt{1-\frac{\alpha}{r_{ij}}} \left| {}_A\!\!\scalar{i}{\psi_0(t)}\right|^2 \Bigg)\label{21}
	\end{eqnarray}
	where again \Eqref{psi0} can be substituted.

\begin{figure}
		\includegraphics[width=\linewidth]{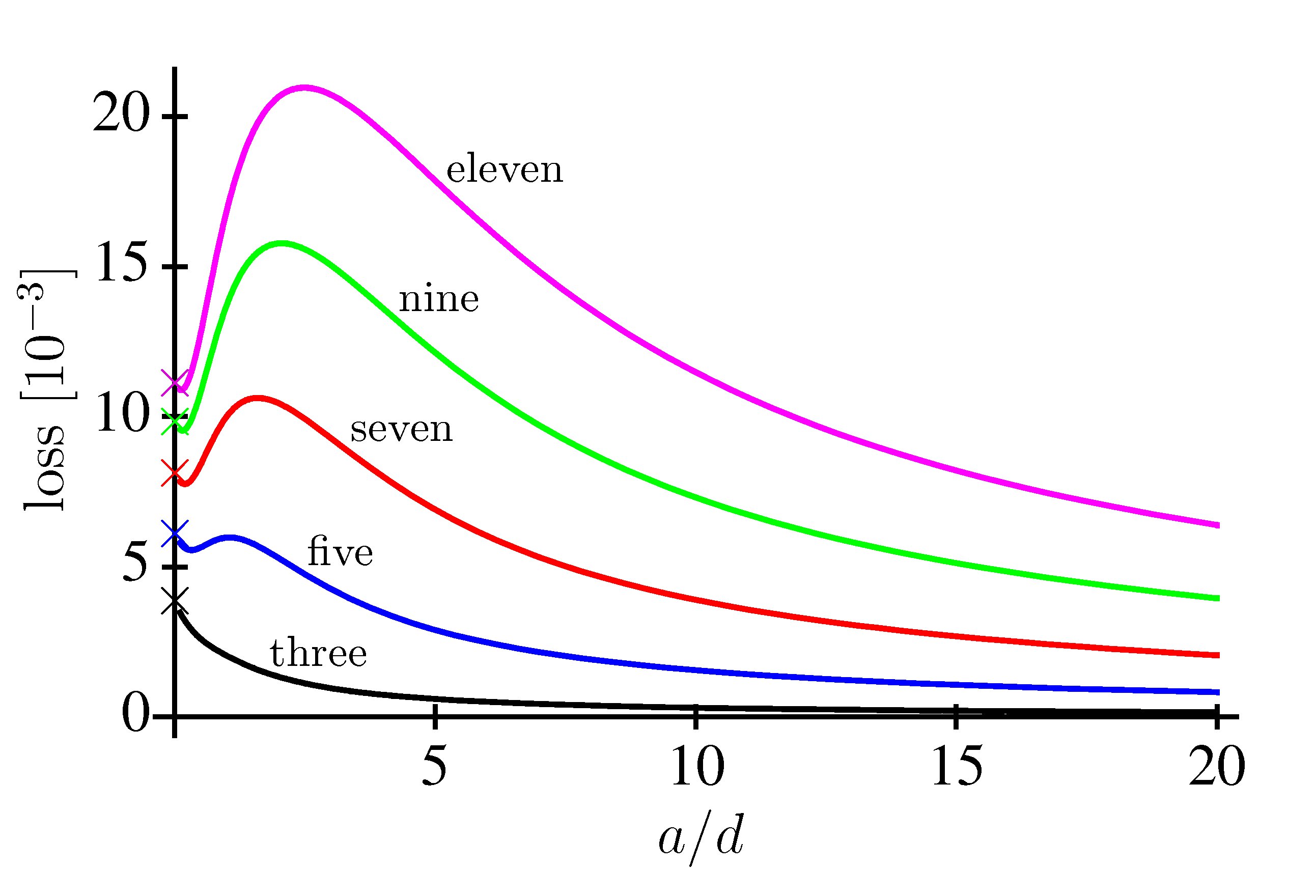}
		\put(-205.8,65){\color{blue}o}
		\put(-200.9,91.1){\color{red}o}
		\put(-196,120){\color{green}o}
		\put(-191.2,149.2){\color{purple}o}\vspace{-2mm}
		\caption{Transfer loss during the transport of an electron along three (black), five (blue), seven (red), nine (green), and eleven (purple) QDs as a function of $\frac{a}{d}$, i.e. the localness of the measurements. The crosses show the solution from the exact \Eqref{4} for local measurments and the circles show the empirical cross-over between local and non-local measurements (see text). Pump and Stokes pulses are as in \Eqref{pulses} (see also Fig~\ref{fig3}(a)) with $T$ chosen such that the additional loss due to non-adiabaticity is $24\times 10^{-6}$ for all curves, to allow reasonable comparison. This is $T=150,\;196,\;225,\;242,\;249$ for three, five, seven, nine, and eleven QDs, respectively. The intermediate couplings are equal one. System parameters are $R=N,\;\alpha=0.04a$. \label{fig4}}
	\end{figure}
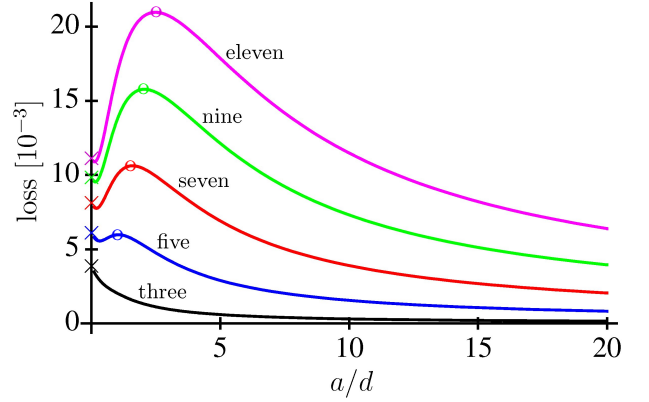

	The terms in the parenthesis of \Eqref{19} and \Eqref{21} become small when the measurements are sufficiently non-local that they can not distinguish well between the QDs involved in the transport (see also Fig~\ref{fig4}). Since the loss of information increases linearly in the time $t_{max}$ required for transport, it is crucial that the couplings $\Omega_j$ between the dots are as large as experimentally possible to give a big energy splitting which in turn allows a fast transport (see \Eqref{15}). 
	
	Also note that in the case of local measurements \Eqref{20} reduces to $\frac{\rm{d}}{{\rm{d} }t}\tilde{\rho}^A_{k0}=-T_2\tilde{\rho}^A_{k0}$ where the decoherence rate $T_2=\frac{R}{N} \left(2+\alpha-2\sqrt{1+\alpha}\right)$ is the same as the one obtained in~\Eqref{aaaa} without transport. Therefore, the decoherence rate of a charge qubit on a quantum dot rail is the same during storage as during transport by CTAP if subject to local measurements. 	
	
	The probability of not finding the electron in the desired state $\ket{n}_A$ after the transport, calculated from \Eqref{19} is shown in Fig~\ref{fig4} as a function of the parameter $\frac{a}{d}$ for transport along 3, 5, 7, 9, and 11 dots. {\color{black}Remember that $\frac{a}{d}$ is the distance of the QPCs to their nearest QD compared to the distance between neighboring QDs, and therefore is a measure of the non-localness of the measurements performed by the QPCs. For any given $a$ this parameter can be changed with $d$ and we fix the sensitivity $\phantom{a}\bra{i}_{\hspace{-6mm}A}\hspace{4mm}\pi_i\ket{i}_A$ of the measurements by setting $\alpha=0.04a$. Also the rail of dots and measurement apparatuses extends on both sides ($N\gg m-n$).
	
	 One can distinguish two regimes in Fig~\ref{fig4}. First, for $\frac{a}{d}>\frac{n-m}{4}$ (right of the circles) the transport fidelity increases with the non-localness of the measurements. This is easily understood since sufficiently non-local measurements can not distinguish between the QDs involved in the transport.
	 
	 Second, for $\frac{a}{d}<\frac{n-m}{4}$ (left of the circles) we find the reverse and the transport fidelity increases with the localness of the measurements (except for transport along three QDs). This needs some explanation. In this regime the measurements are local enough to distinguish well between the states $\ket{m}_A$ and $\ket{n}_A$ (during CTAP most of the population is found on these two states) and hence one can not expect a further decrease of the fidelity with the localness of the measurements. To account for the apparent increase of the fidelity with localness we need another argument. 
	 For very local measurements, sites $n$ and $m$ almost exclusively contribute to the decoherence whereas if the measurements are slightly non-local, sites in the near neighborhood of $n$ and $m$ also contribute to decoherence, therefore increasing the over all decoherence.
	 
	 The small decrease of the fidelity for extremely local measurements might be due to the small populations on QDs other than $m$ and $n$.


	 It is also evident from Fig~\ref{fig4}, that if local dephasing is the main source of decoherence, it is best  to transport long distances in one step as is seen e.g. for the transfer $\ket{1}_A\rightarrow\ket{11}_A$. By doing so the chance of not finding the electron on the desired state after the transport is about $11*10^{-3}$ (see left end of  Fig~\ref{fig4}). This number increases to about $5*(4*10^{-3})$ when we first transfer to $\ket{3}_A$, then to $\ket{5}_A$, $\ket{7}_A$, $\ket{9}_A$, and finally to $\ket{11}_A$. On the contrary, if dephasing tends to be more global, a better transfer rate is achieved by breaking a long distance into several smaller ones.

\subsection{Coupling to TLSs}

	We now apply CTAP to QDs coupled to TLSs as shown in Fig~\ref{fig2}. Hence the driving field $H_A$ from \Eqref{14} has to be added to $H_{int}$ from \Eqref{inter}, responsible for the QD-TLS-coupling. We first neglect the coupling to the measurement apparatus and point out it's inclusion at the end of the subsection. With this all couplings behave local and we can restrict ourself to study only the QDs of the rail which are involved in CTAP, i.e. $N=n-m+1$.
	
	To understand how CTAP works for this system, we note that the Hamiltonian $H=H_{int} +H_A$ can be written in a block diagonal form, with $2^N$ blocks,  each block having dimension $N$ \footnote{A Hamiltonian $H=H_A+H_{TLS}+H_{int}$ with $[H_{TLS},H_{int}]=0$ always allows such a block diagonal form.}. Each block gives the Hamiltonian of the electron on the rail of QDs conditioned on the state of the TLSs. As an example we show part of the Hamiltonian for $N=3$ explicitly 
	\begin{eqnarray}
		\hspace{-0mm} H= \left(  \begin{array}{cccccccccc}
			\!-\chi_1\! & \Omega_1 & 0 \\
			\Omega_1 & -\chi_2 & \Omega_2&&0&&&0&&\cdots \\
			0 & \Omega_2 & \!-\chi_3 \!\\
			&&& \!\!-\chi_1\!\! & \Omega_1 & 0 \\
			&0&&\Omega_1 & -\chi_2 & \Omega_2 &&0&&\cdots\\
			&&&0 & \Omega_2 & \chi_3 \\
			&&&&&&\!-\chi_1 & \Omega_1 & 0 &\cdots\\
			&0&&&0&&\Omega_1 & \chi_2 & \Omega_2 &\cdots\\
			&&&&&&0 & \Omega_2 & -\chi_3\! &\cdots\\  
			&\vdots&&&\vdots&&\vdots&\vdots&\vdots&\ddots       \end{array} \right) \nn 
	\end{eqnarray}
 	where the first block is for all TLSs in the $\ket{0}$ state, the second for the first and second TLSs in $\ket{0}$ and the third TLS in $\ket{1}$, and so on. 
		
	We can now study each block individually because there is no coupling between them. If we want to transport the electron from the first QD to the third QD, regardless in which state or superposition the TLSs are, we have to make sure that CTAP works in each block. That is the state $\ket{1}_A\ket{0}\ket{0}\ket{0}$ should adiabatically evolve to $\ket{3}_A\ket{0}\ket{0}\ket{0}$ as well as $\ket{1}_A\ket{0}\ket{0}\ket{1}$ should evolve to $\ket{3}_A\ket{0}\ket{0}\ket{1}$ and so on. If this is achieved, then an electron initially found in $\ket{1}_A$ will evolve in the pure state $\ket{3}_A$. During the transition the reduced state $\rho^A=$Tr$_{TLSs}[\rho]$ is not only in a superposition of $\ket{1}_A$, $\ket{2}_A$ and $\ket{3}_A$, but in a real mixture of these states (see Fig~\ref{mixture}(a)). This is because depending on the state of the TLSs, the electron might start moving earlier or later, which generally results in entaglement between the rail of QDs and the TLSs. However, once CTAP is completed, the electron will be found in the desired state $\ket{3}_A$.
	
	Hence, for CTAP to be applicable despite coupling to TLSs, we only have to make sure that it works for a Hamiltonian of the form (where we return to an arbitrary long rail of dots)
	\begin{eqnarray}
		\hspace{-0mm}H = \left( \begin{array}{cccccc} \vspace{1mm}
						 \pm \chi_m & \Omega_m&0&\cdots & 0&0  \\ \vspace{-1mm}
			\Omega_m & \pm \chi_{m+1} &\Omega_{m+1} &  \cdots &0& 0 \\ 
			 0 & \Omega_{m+1} & \pm\chi_{m+2} &\ddots &0&0 \\
			\vdots &  \vdots &\ddots &\ddots & \ddots& \vdots  \\ \vspace{1mm}
			 0 &0&0 &\ddots& \pm\chi_{n-1} & \Omega_{n-1}  \! \\
			 0 & 0 &0&\cdots&  \Omega_{n-1}  & \!\pm\chi_{n}\!   			
		\end{array} \right) \label{12345}
	\end{eqnarray}
	instead of the simpler form \Eqref{14}. CTAP should work for all possible permutations of + and $-$, since each permutation represents one block of the Hamiltonian corresponding to a certain combination of environmental TLS states. The crucial requirement for this is again \Eqref{15}, but this time it can not be fulfilled by just performing the transport sufficiently slowly. One also has to make sure, that no level crossings occur, i.e. $\varepsilon_0(t)-\varepsilon_j(t)\ne0$ during step two. As before, $\varepsilon_0(t)$ is the energy of the state $\ket{\psi_0(t)}$ which we want to evolve adiabatically from $\ket{\psi_0(t_0)}=\ket{m}_A$ to $\ket{\psi_0(t_{max})}=\ket{n}_A$. Note that unlike in the previous subsection, $\varepsilon_0(t)\ne 0$ and depends on the state of the TLSs. 
	The situation becomes even more complicated when anti-crossings are considered which appear quite naturally (see Fig~\ref{fig3}). At an anti crossing two energy levels get so close that adiabaticity can not be achieved with any reasonable transfer times.
	
	Furthermore it might happen that the adiabatic state $\ket{\psi_0(t)}$ does not connect $\ket{m}_A$ to $\ket{n}_A$, but instead to another state $\ket{\psi_0(t_{max})}=\ket{j}_A$ with $j\neq n$. An example of this is shown in Fig~\ref{fig3} (blue curves) where the population returns to the original QD.
	
	To calculate the energies is generally only possible numerically and an exact treatment when level crossings or anti-crossings appear is highly non-trivial. Some work on this issue is done in~\cite{multilevel}, but the possibility of anti-crossings is not considered there (see also~\cite{detuning}). However, some qualitative statements can be said. If the energy level $\varepsilon_0(t)$ of the adiabatic state is at all times during step two in the center of all energy levels involved in the transport, i.e.
	\begin{eqnarray}
		\varepsilon_0(t)>\varepsilon_j(t) &&\mbox{for}\quad j=\frac{m-n}{2},\dots ,-1\nn\\
		\varepsilon_0(t)<\varepsilon_j(t) &&\mbox{for}\quad j=1,\dots,\frac{n-m}{2},\label{cond}
	\end{eqnarray}
	then we can be sure that no level crossings occur. In this case the process during  step two is qualitatively the same as without coupling to the TLSs. Since condition~(\ref{cond}) is valid with $\chi_j=0$ (previous subsection), it can always be achieved with sufficiently high driving fields $\Omega_j$, as then the comparatively small couplings $\chi_j$ do not influence the qualitative eigenvalue structure. 
	
	 Lets apply this condition to the block of the Hamiltonian in which we have minus for $\chi_m$ and pluses for all other coupling constants. Clearly we need some strong couplings $\Omega_j(t_0)$ to ensure that $\varepsilon_0(t_0)=-\chi_m$ is not the lowest energy, but the centrefold one (see Fig~\ref{fig3}(c) where this crossing occurs at $t\approx 5$ before the transport starts at $t_0\approx 20$). For $n-m=2$ we find $\Omega_{n-1}^2(t_0)>(\chi_1-\chi_3)(\chi_1-\chi_2)$. For transport along many QDs ($n-m\gg 2$) we get the approximate inequality $\Omega_j(t_0) \gtrapprox c\max(\chi_k)(n-m)$ for some constant $c$, which states that for given maximal $\Omega_j$ (due to the experimental setup) and given $\chi_j$, there exists a maximum number $n-m$ for which we can use CTAP. For transport along more QDs one has to break up the transport into several smaller sections and apply CTAP successively to each section. The shorter the rail gets (compared to the maximal length), the further we can stay away from any level crossing and the faster we can transport the electron without violating the adiabaticity condition~(\ref{15}).

	\begin{figure}
		\centering
		\vspace{-0mm}
		\parbox{4mm}{{\footnotesize(a)}\vspace{68mm}\\\phantom{a}}\hspace{-3mm} \parbox{36mm}{\vspace{-35mm}\includegraphics[width=3.7cm]{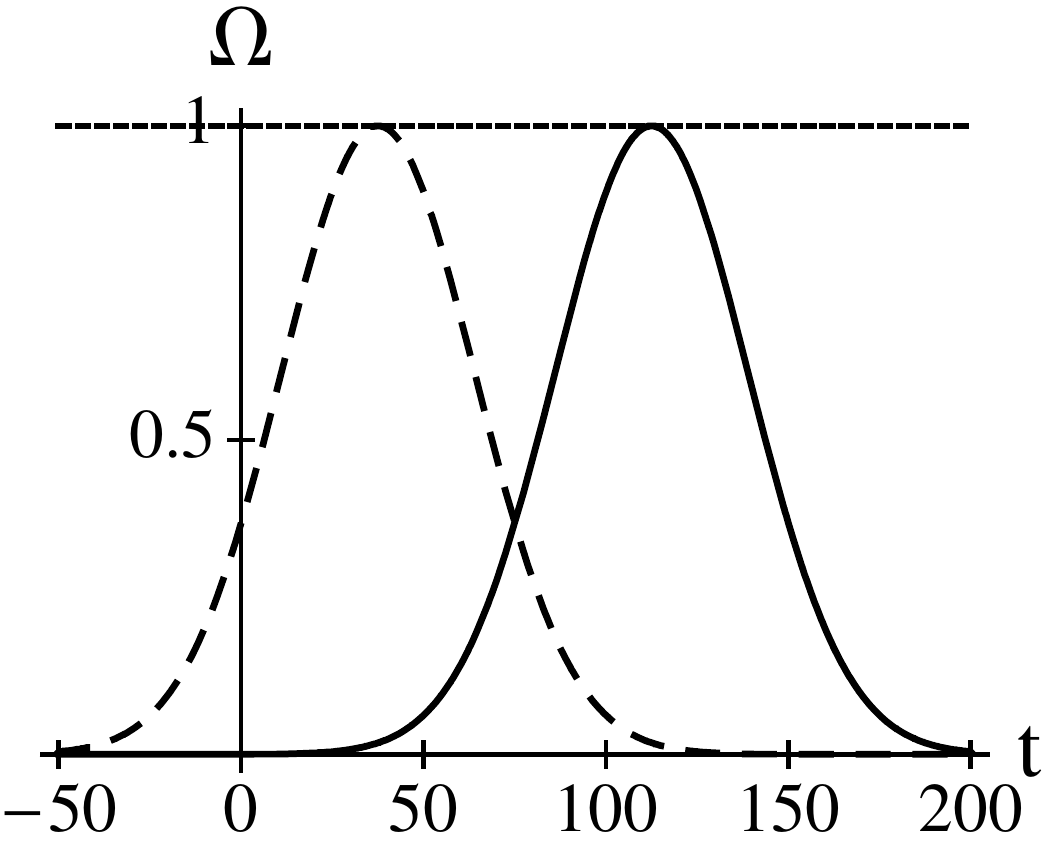}}
		\hspace{-1mm}\parbox{4mm}{{\footnotesize(f)}\vspace{68mm}\\\phantom{a}}\includegraphics[width=4.3cm]{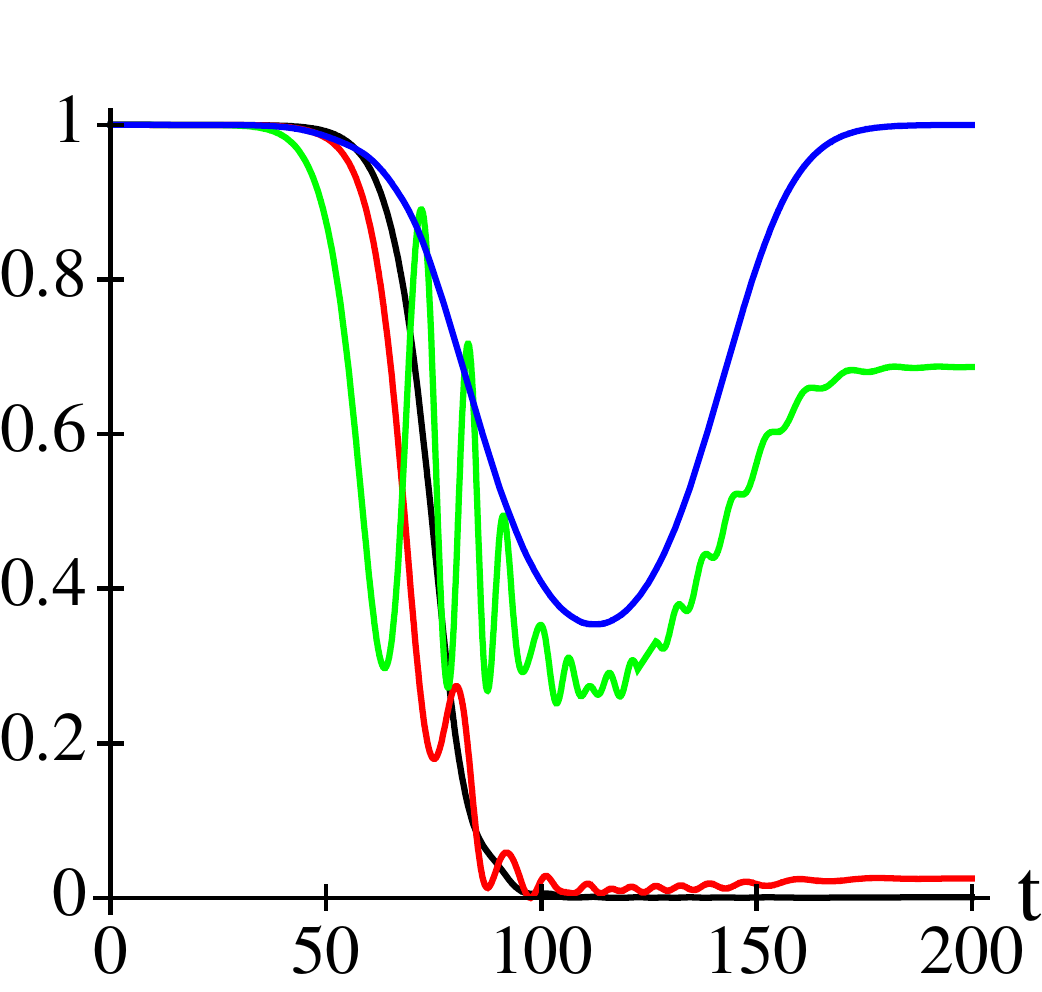}
		\put(-115,109){\normalsize $\mathbf{\rho_{11}}$}\\
		\vspace{-36mm}
		\parbox{4mm}{{\footnotesize(b)}\vspace{68mm}\\\phantom{a}}\hspace{-3mm} \parbox{36mm}{\vspace{-35mm}\includegraphics[width=3.7cm]{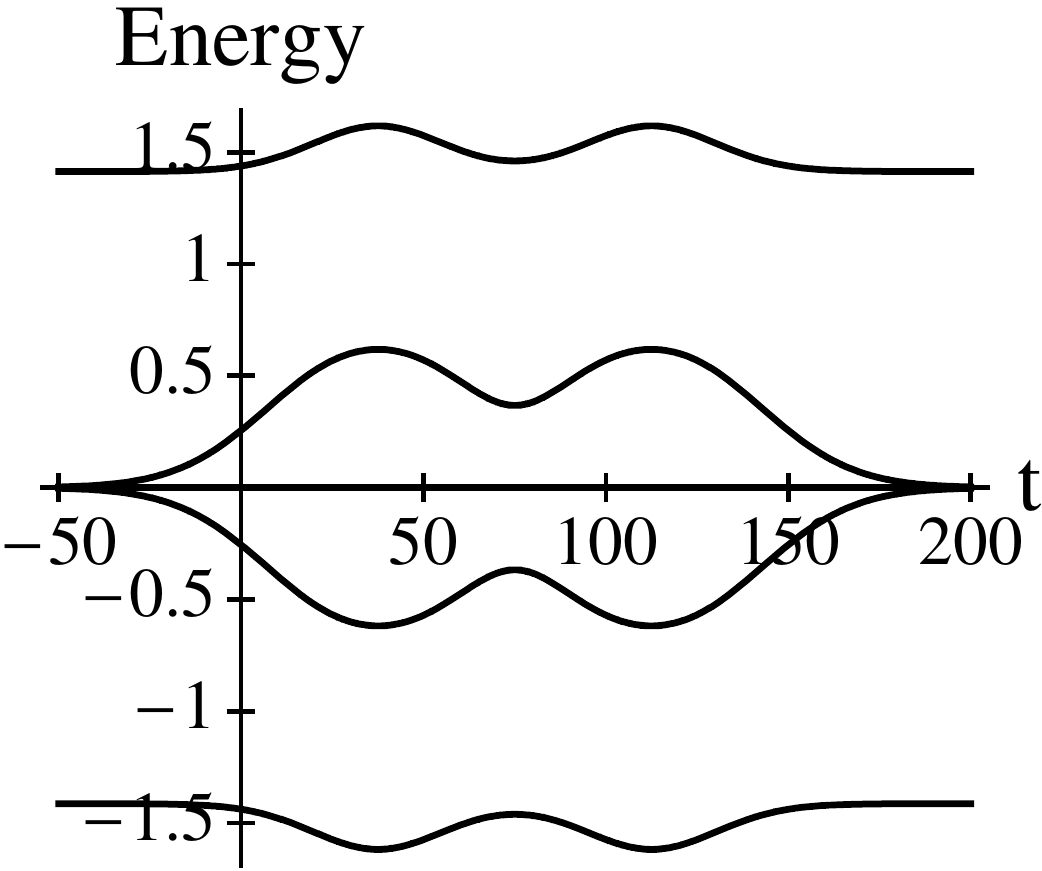}}
		\hspace{-1mm}\parbox{4mm}{{\footnotesize(g)}\vspace{66mm}\\\phantom{a}}\includegraphics[width=4.3cm]{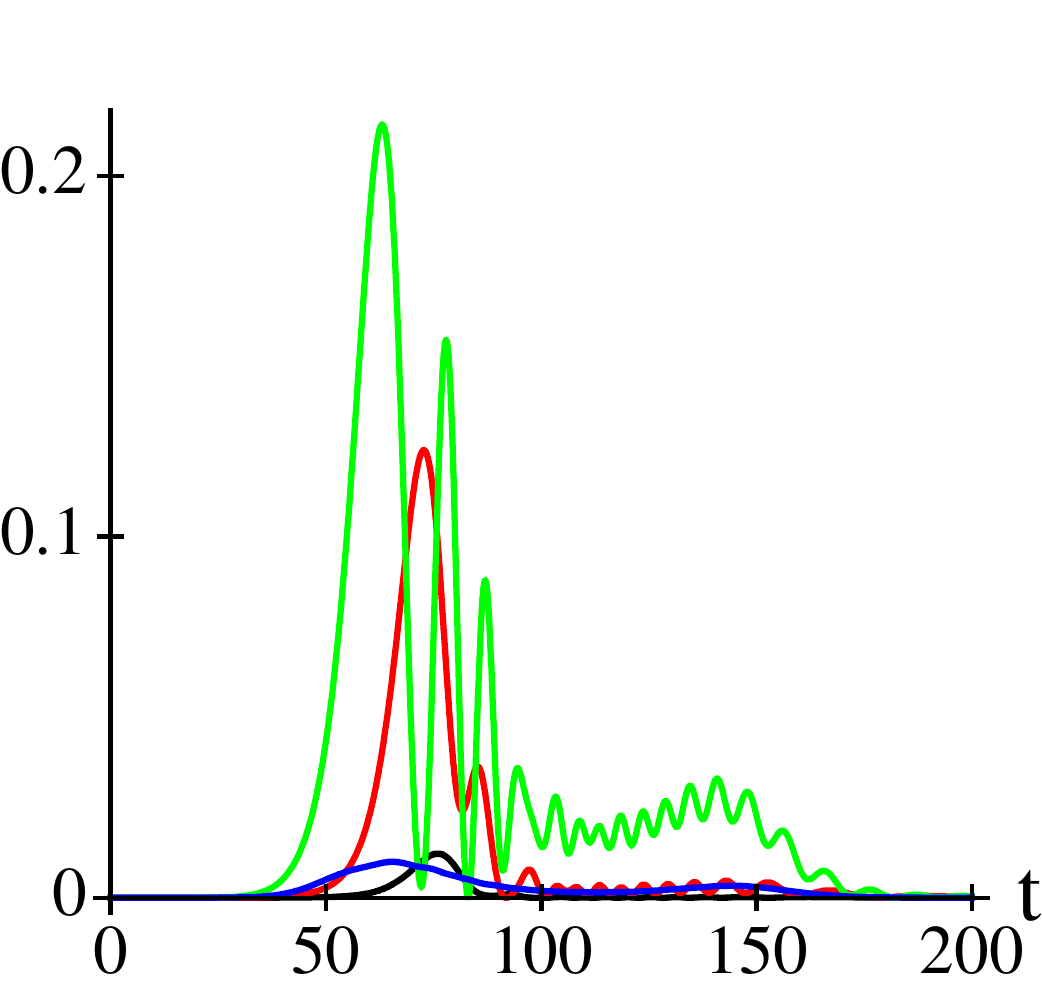}
		\put(-115,108){\normalsize $\mathbf{\rho_{22}}$}\\
		\vspace{-36mm}
		\parbox{4mm}{{\footnotesize(c)}\vspace{68mm}\\\phantom{a}}\hspace{-3mm} \parbox{36mm}{\vspace{-35mm}\includegraphics[width=3.7cm]{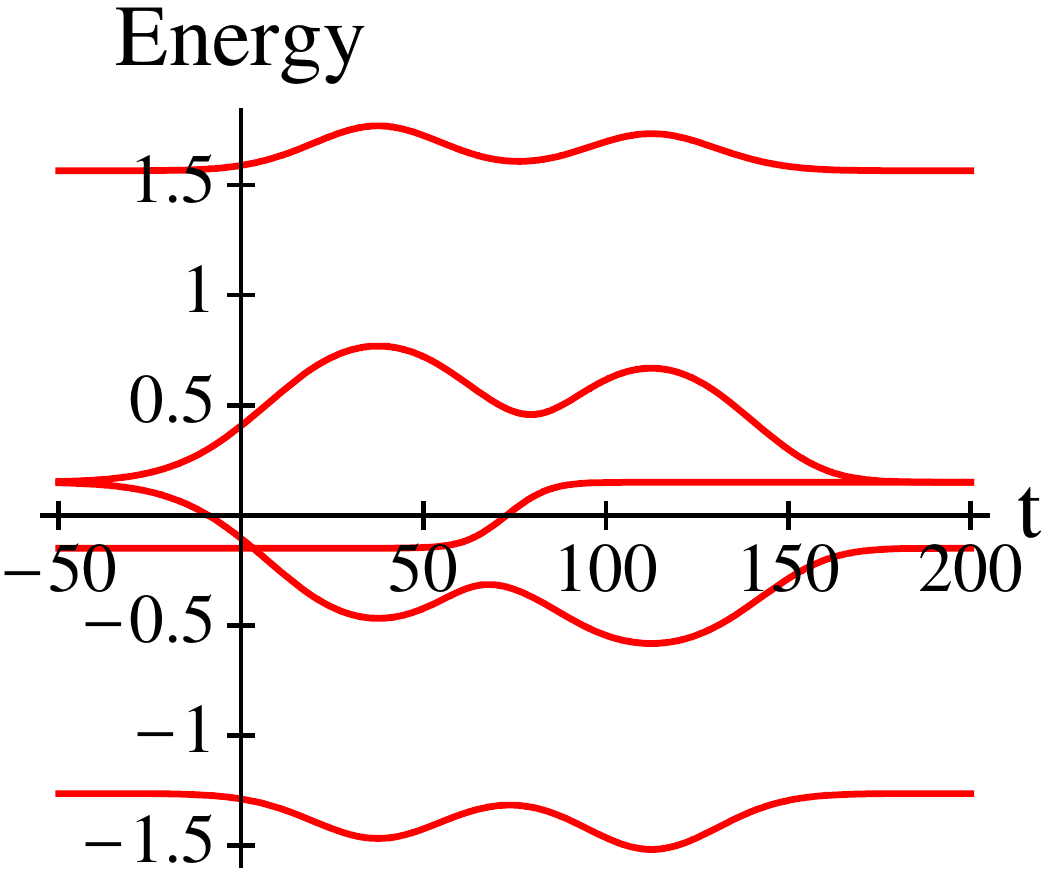}}
		\hspace{-1mm}\parbox{4mm}{{\footnotesize(h)}\vspace{68mm}\\\phantom{a}}\includegraphics[width=4.3cm]{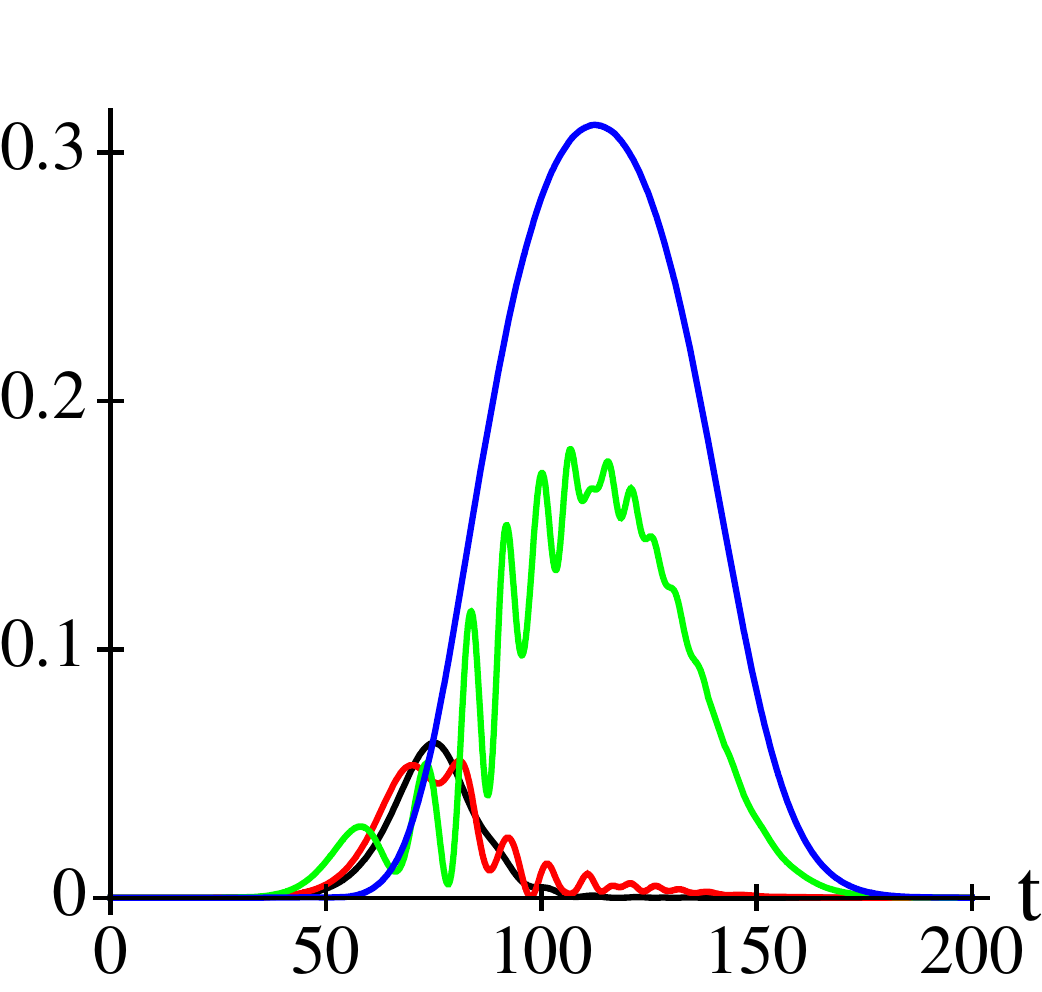}
		\put(-115,108){\normalsize $\mathbf{\rho_{33}}$}\\
		\vspace{-36mm}
		\parbox{4mm}{{\footnotesize(d)}\vspace{68mm}\\\phantom{a}}\hspace{-3mm} \parbox{36mm}{\vspace{-35mm}\includegraphics[width=3.7cm]{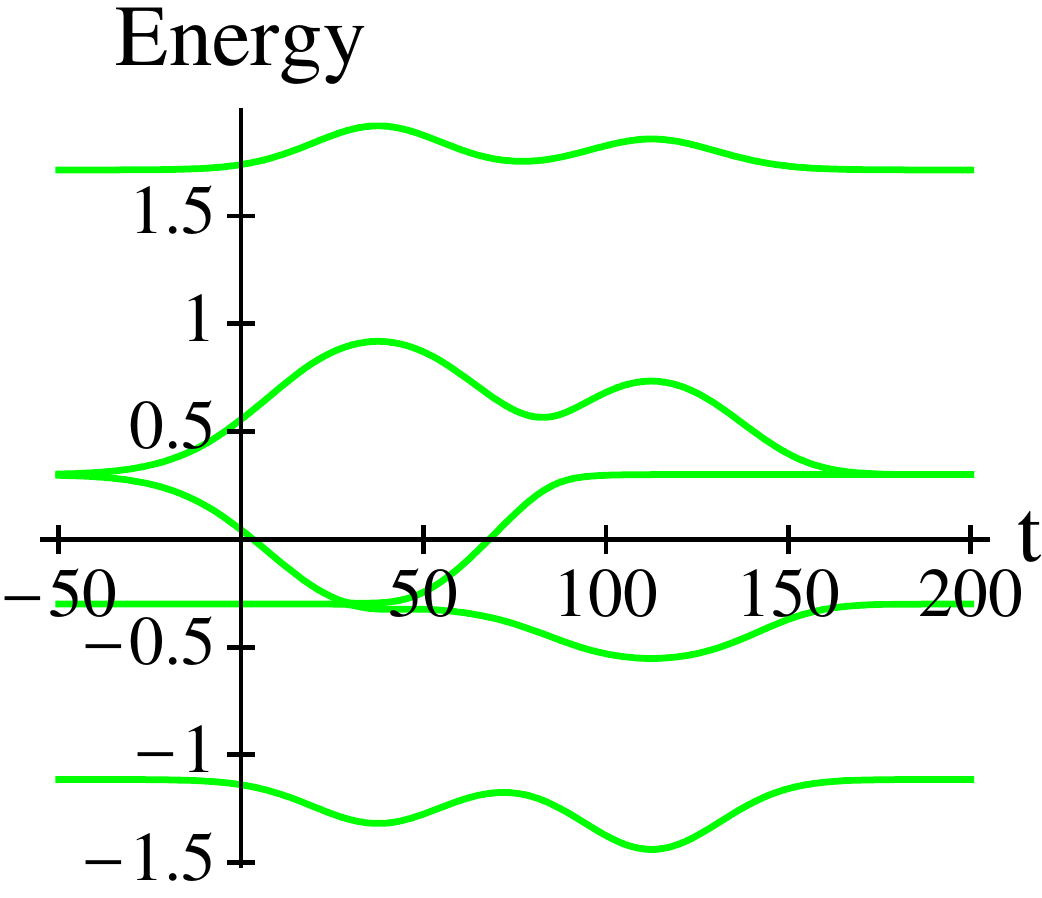}}
		\hspace{-1mm}\parbox{4mm}{{\footnotesize(i)}\vspace{68mm}\\\phantom{a}}\includegraphics[width=4.3cm]{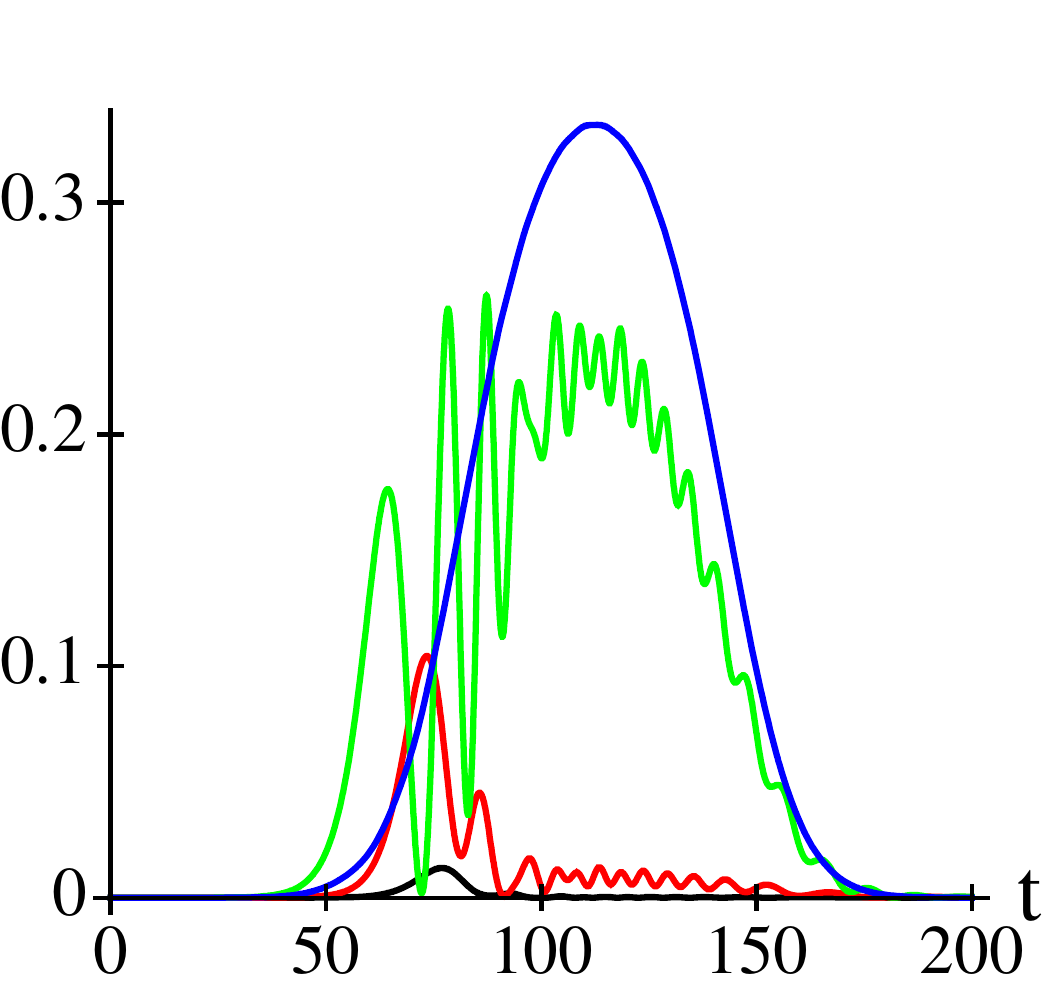}
		\put(-115,108){\normalsize$\mathbf{\rho_{44}}$}\\
		\vspace{-35.5mm}
		\parbox{4mm}{{\footnotesize(e)}\vspace{68mm}\\\phantom{a}}\hspace{-3mm} \parbox{36mm}{\vspace{-35mm}\includegraphics[width=3.7cm]{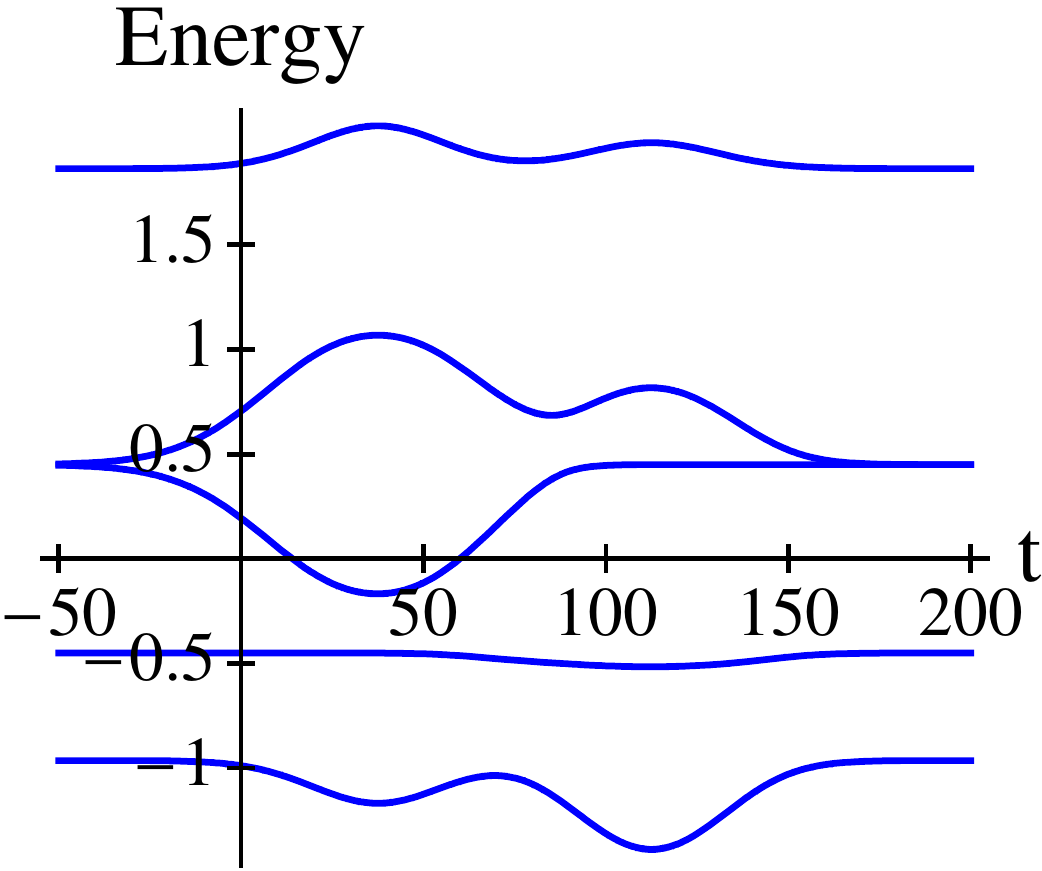}}
		\hspace{-1mm}\parbox{4mm}{{\footnotesize(j)}\vspace{68mm}\\\phantom{a}}\includegraphics[width=4.3cm]{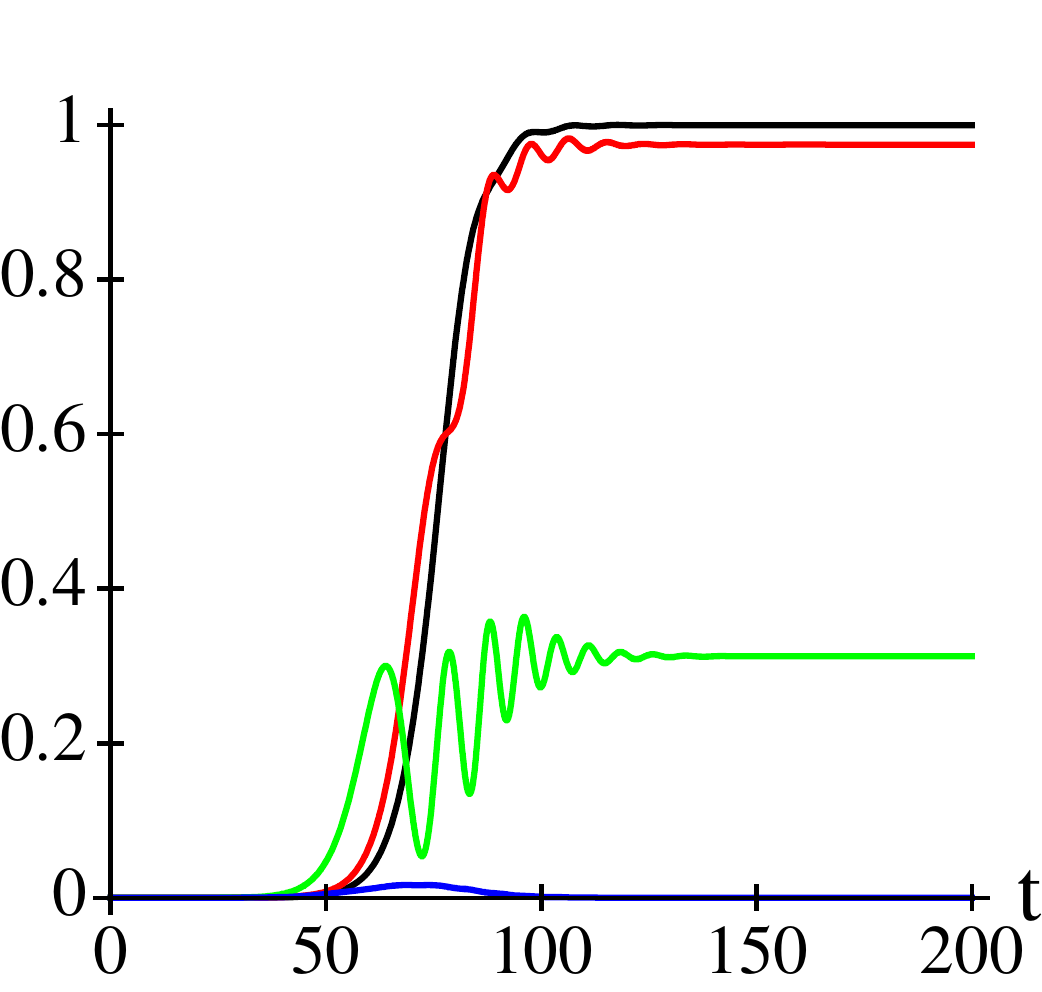}\vspace{-35.5mm}
		\put(-115,109){\normalsize $\mathbf{\rho_{55}}$}
		\caption{(a): Coupling between QDs (dashed:~$\Omega_1$, {\color{black}dotted:~$\Omega_{2,3}$, solid:~$\Omega_4$}), (b)-(e): energy levels of Hamiltonian~\Eqref{Hami}~and $\chi=0,\;0.15,\;0.3,\;0.45$ (see text), and (f)-(j): populations of $\ket{1}_A$ -- $\ket{5}_A$ obtained by the numerical integration of the Schr\"odinger equation. The colors black, red, green and blue correspond to $\chi=0,\;0.15,\;0.3,\;0.45$, respectively.~What~looks like level crossings in (c) and (d) are actually anti-crossings.  \label{fig3}}
	\end{figure}
	
	As example we show in Fig~\ref{fig3} the eigenenergies and populations as a function of time for a  system with five QDs and constant $\chi_i=\chi$ for different values of $\chi$. We consider the block of the Hamiltonian acting in the subspace where the first TLS is in the state $\ket{0}$ and the other four TLSs are in the state $\ket{1}$, which reads
	 \begin{eqnarray}
		\hspace{-0mm}H_A = \left( \begin{array}{ccccc} 
			- \chi & \Omega_P&0 & 0&0  \\ 
			\Omega_P & + \chi &\Omega_{2}   &0& 0 \\ 
			 0 & \Omega_{2}  & +\chi  &\Omega_3&0 \\
			 0 &0&\Omega_3 & +\chi & \Omega_{S}   \\
			 0 & 0 &0&  \Omega_{S}  & +\chi  			
		\end{array} \right). \label{Hami}
	\end{eqnarray}
	Of course, if we assume the TLSs to be in a statistical mixture one would have to calculate the populations for all blocks of  the Hamiltonian and average over them. Since all qualitative features can be seen with this particular block we restrict the discussion to this one for pedagogic reasons. We use $\Omega_2=\Omega_3\equiv 1$ and  Gaussian pump and Stokes pulses 
	 \begin{eqnarray}
	 	\Omega_P &=& \exp\left\{ -\frac{(t-3T/4)^2}{(T/4)^2}   \right\} \nn\\ 
		\Omega_S &=& \exp\left\{ -\frac{(t-T/4)^2}{(T/4)^2}   \right\} \label{pulses}
	\end{eqnarray}
	with the parameter $T=150$. Because of the Gaussian pulse shapes the distinction of steps one, two and three is not sharp, any more.  Looking at Fig~\ref{fig3}~(a) we can approximately say that step one is for $-50<t<20$, step two is for $20<t<130$ and step three  is for $130<t<200$. $T$ is  chosen long enough to achieve very good transfer fidelity (0.9996) for $\chi=0$, i.e. without coupling to TLSs. For $\chi=0.15$ there appears a level anti-crossing at which the energy gap (0.0002) is so small, that in can be treated as a crossing. However, this happens during step one before the actual population transfer starts and does therefore not disturb adiabaticity much.  The smaller population transfer (0.975) is because during step two the energy level spacing is decreased compared to the previous case. Therefore, the transfer rate can be increased arbitrarily close to one by choosing larger $T$ \footnote{Non-adiabatic corrections scale with $\exp{(-cT|\varepsilon_0-\varepsilon_j|)}$ for some constant $c$ (see e.g.~\cite{corrections}).}. Enlarging $\chi=0.3$ moves the anti-crossing to step two and adiabaticity can not be achieved any more. The result is wild oscillations of all populations and a low transfer rate. Finally at $\chi=0.45$ the anti-crossing disappears and adiabaticity is restored which is evident in the lack of oscillations in the populations. But because the eigenenergy -0.15 at $t=-\infty$ connects to the same energy at $t=\infty$, the population returns to the original state $\ket{m}_A$.
		
	This result should be quite surprising. Up to relatively large couplings to TLSs, $\frac{\chi_i}{\Omega_{j,max}}<0.2$ (for $m-n=4$) adiabatic transfer can be achieved and the transfer fidelity {\color{black}very quickly approaches one with sufficiently} large transfer times. This is contrary to the Markovian dephasing studied in the previous subsection, where the transfer loss increases with transfer time once the transfer time is long enough to ensure adiabaticity. 
	 	
	One disturbing effect introduced by the coupling to TLSs is that $\varepsilon_0\ne 0$ and hence we get a dynamical phase 
	\begin{equation}
		\varphi=\int_0^{t_{max}} \dd t\, \varepsilon_0(t)
	\end{equation}
	and worse even, this phase depends highly on the state of the TLSs. If the electron starts in a superposition $\frac{1}{\sqrt{2}}(\ket{k}_A +\ket{m}_A)$ it will not end up in $\frac{1}{\sqrt{2}}\left(\ket{k}_A +e^{\imath\varphi}\ket{n}_A\right)$ but in a real mixture of $\ket{k} _A $ and $\ket{n} _A $. That is, during the transport, the electron gets entangeled with the TLSs. This should not be surprising since we saw a similar behavior already in subsection~\ref{coupling1} without transport (see \Eqref{13}). However, CTAP prevents the lost information from returning back to the electron periodically. This can be seen in Fig~\ref{mixture}(b) for transport along three QDs, $\chi_i\equiv\chi$, and $\Omega_P ,\;\Omega_S$ as in \Eqref{pulses} with $T=150$. The initial states  $\rho_j(t_0)=\half\one$ of the TLSs are taken to be a complete mixture. As a measure of purity we use Tr$\left[\left(\rho^A\right)^2\right]$.
	\begin{figure}
		\centering
		\vspace{-0mm}
		\hspace{-1mm} \parbox{4mm}{{\footnotesize(a)}\vspace{62mm}\\\phantom{a}}\hspace{-1mm} \parbox{40mm}{\vspace{-35mm}\includegraphics[width=3.8cm]{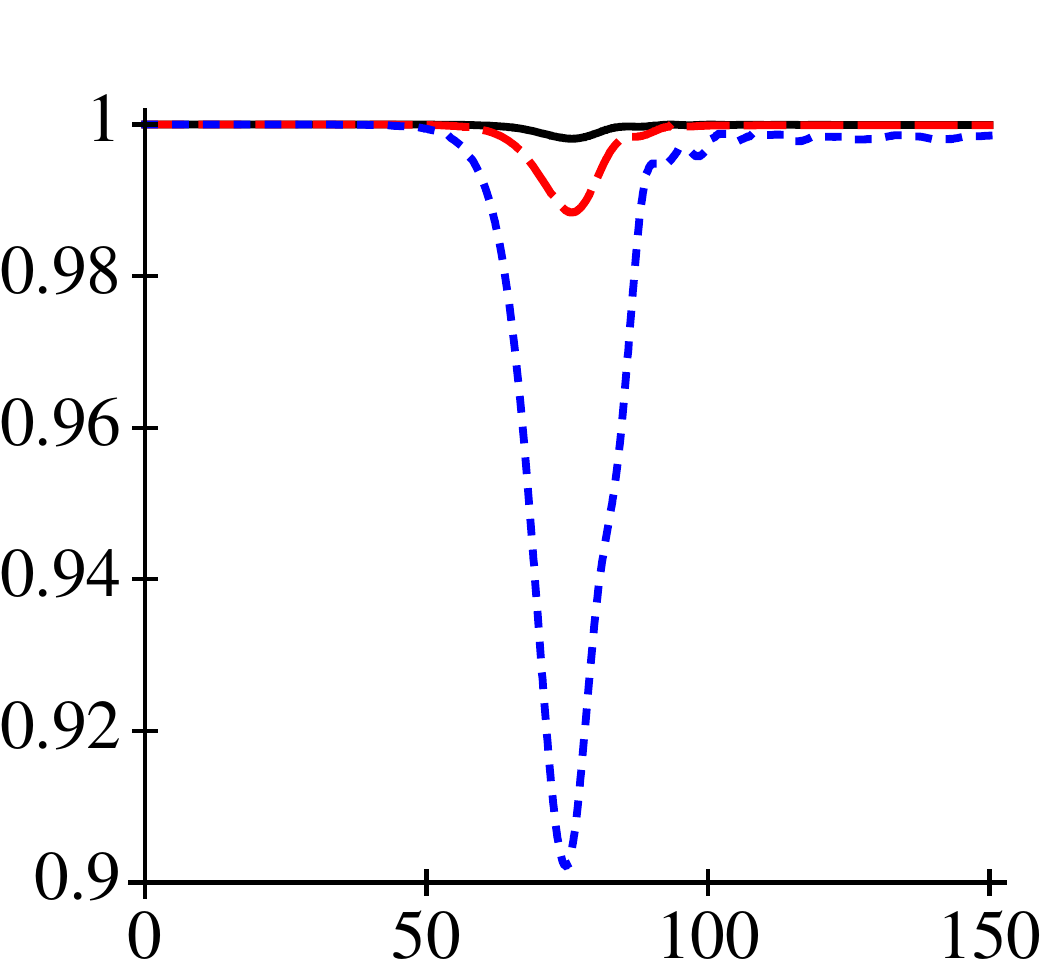}
		 }				
		\hspace{-0.5mm}\parbox{4mm}{{\footnotesize(b)}\vspace{62mm}\\\phantom{a}}\includegraphics[width=3.8cm]{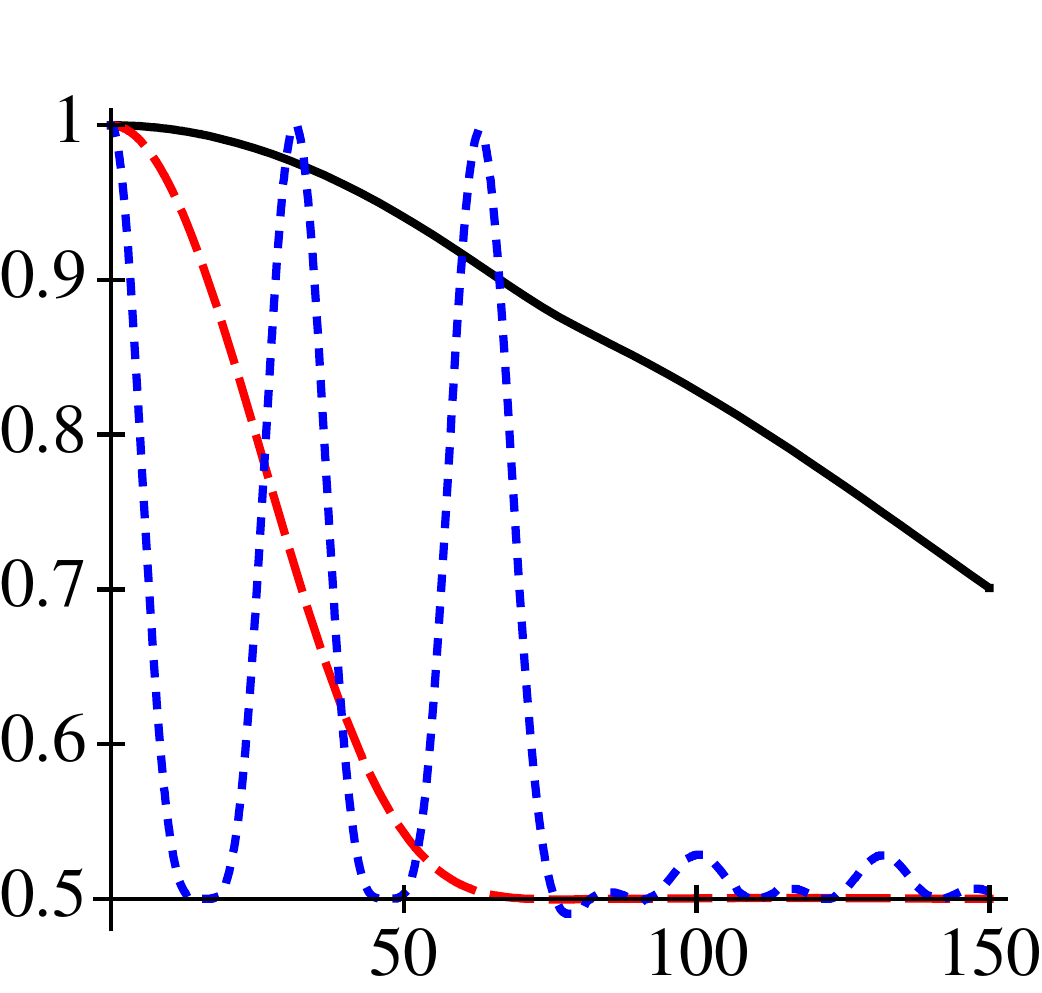}
		\put(-50,-9){t}\put(-175,-9){t}
		\put(-237,32){\begin{rotate}{90}Tr$[(\rho^A)^2]$\end{rotate}}
		\put(-113,32){\begin{rotate}{90}Tr$[(\rho^A)^2]$\end{rotate}}
		\vspace{-30mm}
		\caption{The purity of the reduced density matrix $\rho^A(t)$ during the transfer $\ket{1}_A\to\ket{3}_A$.\\
		(a): Transport of a position eigenstate. $\ket{\psi(t=0)}=\ket{1}_A$ and $\chi_i=0.02$~(solid), 0.05~(dashed), 0.15~(dotted).  Because of the dependance of $\ket{\psi_0(t)}$ on the state of the TLSs, the electron gets entangled with the TLSs during transport. However, this entanglement disappears with the completion of CTAP and the electron on the quantum dot rail finishes the transport in the desired pure state.\\
		(b): Transport of half of a superposition state. $\ket{\psi(t=0)}=\frac{1}{\sqrt{2}}\left(\ket{0}_A+\ket{1}_A\right)$ and $\chi_i=0.005$~(solid), 0.02~(dashed), 0.1~(dotted). The transported half of the superposition state picks up a dynamical phase which depends on the state of the TLSs. Hence the electron gets entangled with the TLSs and finishes CTAP in a statistical mixture.  \label{mixture}}
	\end{figure}
	On the other hand, as described earlier this section, if the electron starts in $\ket{1}_A$ it will get transfered to the pure state $\ket{3}_A$ despite being in a real mixture during the transfer as is shown in Fig~\ref{mixture}(a).
		
	Now we briefly discuss the inclusion of measurements once again. In the previous subsection the adiabatic theorem was assumed to hold and therefore we first have to make sure that it holds also with coupling to TLSs. This is best done by solving the Schr\"odiner equation without measurements as was done to get Fig~\ref{fig3}. The solution shows whether the transformation is adiabatic (almost no fast oscillations) or not (much oscillations). If not, the driving fields $\Omega_j$ as well as the transfer time can be increased, until the solution shows adiabatic behavior. Then equations (\ref{18})-(\ref{21}) derived in the previous subsection can be applied here as well, if one uses the appropriate eigenstate $\ket{\psi_0(t)}$ which depends on the state of the TLSs. However, since $\ket{\psi_0(t)}$ can only be calculated numerically and the effects of dephasing do not depend much on the exact form of this state, one might prefer to use the unperturbed eigenstate from \Eqref{psi0} as in subsection~\ref{hello}. It is therefore justified to examine decoherence effects arising due to measurements and due to coupling to TLSs separately.

\section{Discussion and Conclusion\label{conclusion}}

	While storing information as a charge qubit on a quantum dot rail, some of it will leak because of decoherence due to measurements. Information loss from decoherence arising from coupling to TLSs, on the contrary, returns periodically. 
	
	On the other hand, while transporting information along a rail of QDs with CTAP, information lost to TLSs will not return to the electron on the QDs. However, only information of the phase will be lost since in the adiabatic limit one can achieve perfect population transport despite coupling to TLSs, provided it is not too strong compared to the driving field $\Omega_j$ and the QDs involved are few enough. This is different for decoherence through measurements when there is always a chance of loosing the electron during the transport, which can be minimized by doing the transport fast.
	
	Summarizing we can say, if we can reduce Markovian and non-Markovian dephasing sufficiently to be able to store a charge qubit on a rail of QDs, then we can also transport it using CTAP.
	
	It is often claimed that CTAP is relatively robust to decoherence because the QDs between the initial and the final QDs are barely populated. This is in contrast to our results because of the following reasoning. To achieve a small population on the intermediate QDs the coupling between these has to be large compared to the Stokes and pump pulses, as seen in \Eqref{psi0}. On the other hand we find that all couplings should be as large as possible (for both, Markovian and non-Markovian decoherence) to achieve big energy spacings which in turn allow fast transport. Therefore it is reasonable to choose large Stokes and pump pulses and allow some population on the intermediate QDs. The above statement might stem from analyses of STIRAP where the intermediate atomic states are exposed to decay, but this is not the case for CTAP. 
			
	Although the non-Markovian noise studied here is fairly simple and might not be very realistic in an experiment, the striking differences we find compared to Markovian dephasing indicates that it might be worth studying more specific examples of this class of decoherence.
		
	Another approach to quantum computing is the use of the spin degree of freedom of an electron as a qubit. In this case the differences between the influence of Markovian and non-Markovian dephasing on transport with CTAP might become even more apparent. Because the coupling of the spin degree to the environment is expected to be much weaker than the coupling of the charge, the dynamic phase acquired during the transport due to coupling to TLS's should be fairly independent of the spin and therefore an undetectable global phase even for superposition of spin states.

\end{document}